\documentclass[conference]{IEEEtran}
\IEEEoverridecommandlockouts

\usepackage{color}
\usepackage{bm}
\usepackage{bbm}
\usepackage{cite}
\usepackage{amsmath}
\usepackage{amssymb}
\usepackage{amsthm}
\usepackage{amsfonts}
\usepackage{algorithmic}
\usepackage{graphicx}
\usepackage{mathrsfs}
\usepackage{empheq}	
\usepackage[mathcal]{euscript}
\usepackage[margin = 2cm]{geometry}
\usepackage{framed}
\usepackage{textcomp}
\usepackage{xcolor}
\usepackage{multirow}
\newtheorem{theorem}{Theorem}
\newtheorem{lemma}{Lemma}

\newtheorem{remark}{Remark}
\newtheorem{assumption}{Assumption}

\renewcommand{\P}{\mathbb{P}}
\newcommand{\E}{\mathbb{E}}
\newcommand{\VAR}{{\sf VAR}}
\newcommand{\beq}{\begin{equation}}
\newcommand{\eeq}{\end{equation}}
\newcommand{\beqa}{\begin{eqnarray}}
\newcommand{\eeqa}{\end{eqnarray}}

\newcommand{\thetatrue}{\theta_0}
\newcommand{\T}{^\top}

\newcommand{\mnet}{{\sf m}_{\mathrm{ave}}}
\newcommand{\cnet}{{\sf c}_{\mathrm{ave}}}
\newcommand{\Cnet}{{\sf C}_{\mathrm{ave}}}

\newcommand{\xnet}{\bm{x}_{\mathrm{ave}}}
\DeclareMathOperator*{\argmax}{arg\,max}
\def\BibTeX{{\rm B\kern-.05em{\sc i\kern-.025em b}\kern-.08em
		T\kern-.1667em\lower.7ex\hbox{E}\kern-.125emX}}
\allowdisplaybreaks

\newcommand*{\QED}{\hfill\ensuremath{\blacksquare}}%

\begin{document}

\title{Adaptation in Online Social Learning
\thanks{This work was supported in part by  grant 205121-184999 from the Swiss National Science Foundation (SNSF).}
}

\author{\IEEEauthorblockN{Virginia Bordignon}
\IEEEauthorblockA{\textit{School of Engineering, EPFL} \\
virginia.bordignon@epfl.ch}
\and
\IEEEauthorblockN{Vincenzo Matta}
\IEEEauthorblockA{\textit{DIEM, University of Salerno} \\
vmatta@unisa.it}
\and
\IEEEauthorblockN{Ali H. Sayed}
\IEEEauthorblockA{\textit{School of Engineering, EPFL} \\
ali.sayed@epfl.ch}

}

\maketitle

\begin{abstract}
This work studies social learning under non-stationary conditions. Although designed for online inference, classic social learning algorithms perform poorly under drifting conditions. To mitigate this drawback, we propose the Adaptive Social Learning (ASL) strategy. This strategy leverages an {\em adaptive Bayesian update}, where the adaptation degree can be modulated by tuning a suitable step-size parameter. The learning performance of the ASL algorithm is examined by means of a steady-state analysis. It is shown that, under the regime of small step-sizes: $i)$ {\em consistent learning} is possible; $ii)$ an accurate prediction of the performance can be furnished in terms of a Gaussian approximation.
\end{abstract}

\section{Introduction}
In social learning strategies, a set of communicating agents seeks to update their opinions as they receive streaming information about a given observed phenomenon~\cite{ChamleyBook, Jad, PoorSPmag2013,ScaglioneSPmag2013}. In most existing methods in the literature, as time evolves, agents' opinions (or beliefs) tend to concentrate on the true state~\cite{AcemogluOzdaglar2011,Jadbabaie2013,Zhao,Salami,NedicTAC2017,Javidi,MattaSantosSayedICASSP2019,MattaBordignonSantosSayed2019}, often at an exponentially fast rate of convergence. However, such remarkable convergence properties have the collateral effect of hindering adaptation.

\begin{figure}[htb]
	\centering
	\includegraphics[width=3in]{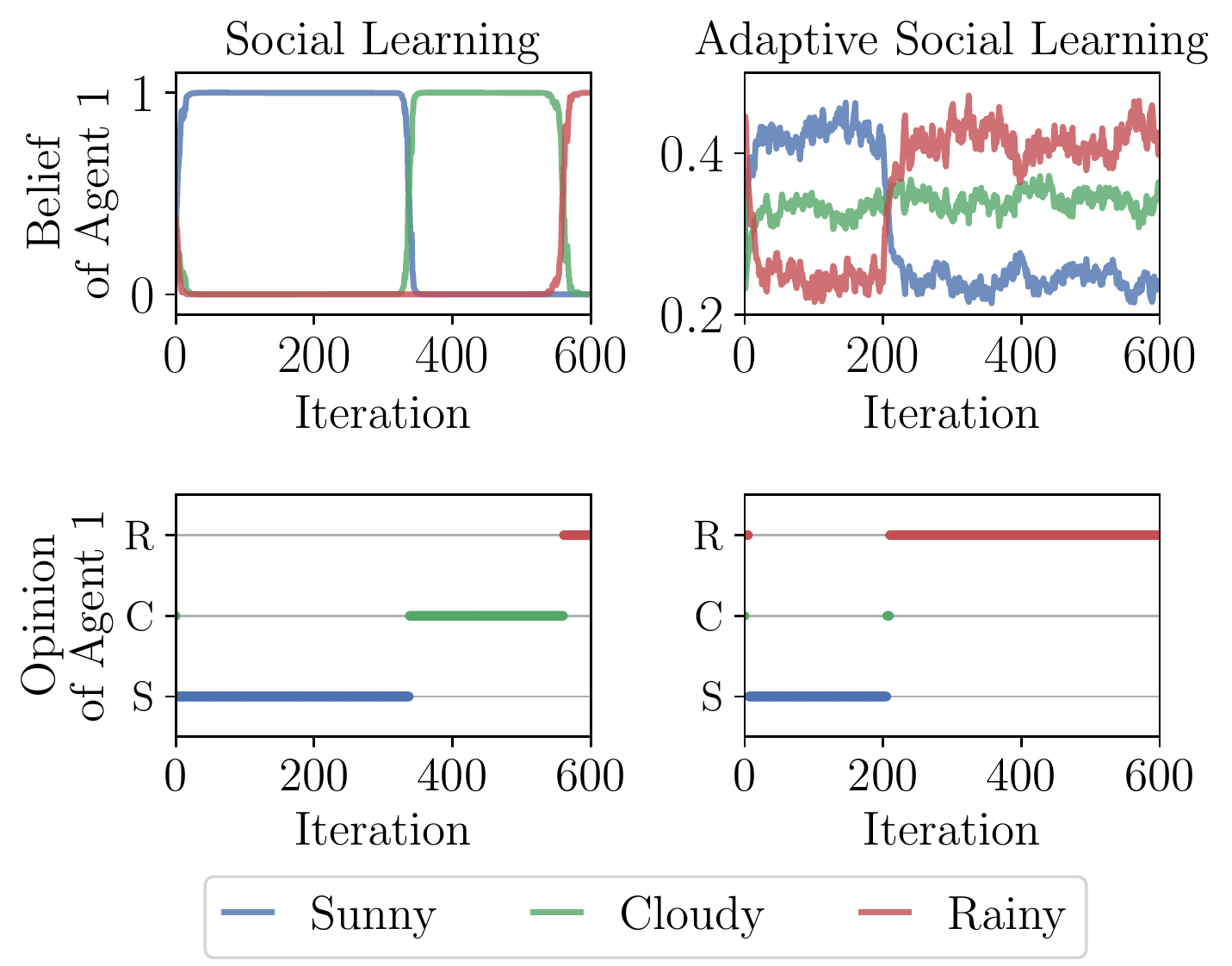}
	\caption{{\em Classic} social learning vs. {\em adaptive} social learning. {\em Top panels}: Belief evolution of agent $1$, with $\thetatrue$ changing at time $i=200$. {\em Bottom panels}: The instantaneous decision taken by agent $1$ by choosing the hypothesis that maximizes the current belief.}
	\label{fig:examplesl}
\end{figure}

Let us consider the following example. A network of $10$ agents aims to solve a weather forecast problem using an online social learning algorithm. At each instant, these agents collect data coming from one among three possible hypotheses: ``{\em sunny}'', ``{\em cloudy}'', ``{\em rainy}''.  At first, data are consistent with the hypothesis ``{\em sunny}'', but then from instant $i=200$ they indicate that the correct forecast is ``{\em rainy}''. As we see in Fig.~\ref{fig:examplesl} (the curves illustrate the behavior of Agent 1), the social learning algorithm reacts with a considerable inertia to the hypothesis drift. 

In fact, Fig.~\ref{fig:examplesl} shows clearly that the agent learns well until instant $i=200$, whereas from $i=200$ onward, the situation changes dramatically: the classic social learning algorithm has a delayed reaction. {First, agents perceive a change only at $i\approx 350$, but start opting for the wrong hypothesis ``{\em cloudy}''. Then, after a prohibitive number of iterations, at $i\approx 550$, agents manage to overcome their stubbornness and opt for the correct hypothesis ``{\em rainy}''}. To tackle this problem, this work proposes an Adaptive Social Learning (ASL) strategy, whose performance is shown in the second column of the same  Fig.~\ref{fig:examplesl} for the same example. We see that the ASL algorithm manages to track the target change at instant $i\approx 200$, exhibiting an adaptation capacity that is remarkably higher than that of the classic social learning algorithm.

The main contributions of this work can be summarized as follows. First, we introduce a novel social learning strategy that enables adaptation. 
Then, by exploiting recent advances in the field of distributed detection over adaptive networks~\cite{MattaSayedCoopGraphSP2018}, we provide an accurate analytical characterization of this strategy in terms of $i)$ convergence of the system at  steady state (Theorem~\ref{theor:steady}); $ii)$ achievability of consistent learning (Theorem~\ref{theor:weaklaw}); $iii)$ a Gaussian approximation for the learning performance (Theorem~\ref{theor:CLT}). Due to space constraints, proofs will be omitted.

\section{ASL Strategy}
Consider a strongly-connected network of $N$ agents trying to infer the true state of nature $\theta_0\in\Theta$ given a set of $H$ hypotheses, $\Theta=\{1,2,\ldots,H\}$. Each agent $k$, at time $i$, observes streaming data $\bm{\xi}_{k,i}$, belonging to a certain space $\mathcal{X}_k$, drawn from a distribution that depends on the underlying hypothesis $\theta_0$. The data are assumed to be independent over time, i.e., across index $i$, whereas they can be dependent across agents. Moreover, it is assumed that the distribution of $\bm{\xi}_{k,i}$ belongs to a set of $H$ admissible models (likelihood functions) that are identified by the hypotheses $\theta\in\Theta=\{1,2,\ldots,H\}$. 
The likelihood of agent $k$ evaluated at $\theta$ is denoted by
$
L_k(\xi|\theta)
$ with $\xi\in\mathcal{X}_k$. Note that the likelihoods are allowed to vary across the agents.

We model the network using a strongly-connected graph, with a left-stochastic combination matrix $A\triangleq[a_{\ell k}]$. Element $a_{\ell k}$ weights information received by agent $k$ from agent $\ell$: $a_{\ell k}$ is a non-negative real number and it is equal to zero if $\ell\notin\mathcal{N}_k$, 
{where $\mathcal{N}_k$ is the neighborhood of agent $k$ ($k$ included).}  
We define the Perron eigenvector $\pi$ such that~\cite{Sayed}:
\beq
A\pi=\pi,\qquad\mathbbm{1}\T\pi=1,\qquad\pi\succ 0.
\eeq

Agents will incorporate the information contained in their local observations and diffuse it across the network by iteratively updating and exchanging their belief vectors $\bm{\mu}_{k,i}$. The belief vector is a probability vector over the set of hypotheses $\Theta$ and each component $\bm{\mu}_{k,i}(\theta)$ reflects the confidence of agent $k$ at instant $i$ that $\theta$ is the true hypothesis.

\begin{assumption}[Positive initial beliefs]\label{assum:initbel}
	All agents start with a strictly positive belief for all hypotheses, i.e., $\bm{\mu}_{k,0}(\theta)>0$ for each agent $k$ and all $\theta\in \Theta$.~\hfill$\square$ 
\end{assumption}

\subsection{Adaptive Social Learning (ASL) Algorithm}
In the adaptive scenario, system conditions can change over time, e.g., the true state of nature or the network topology might change. To address that setup we now introduce the ASL strategy, which can be described in terms of the following iterative two-step algorithm. In the first step, each agent $k$ constructs an {\em intermediate} belief vector $\bm{\psi}_{k,i}$ by incorporating the current observation $\bm{\xi}_{k,i}$ into the belief of the preceding time epoch, $\bm{\mu}_{k,i-1}$, through the following {\em adaptive Bayesian update}: 
\begin{equation}
	\bm{\psi}_{k,i}(\theta)=\displaystyle{
		\frac{\bm{\mu}^{1-\delta}_{k,i-1}(\theta)L^{\delta}_k(\bm{\xi}_{k,i}|\theta)}
		{\sum_{\theta'\in\Theta}\bm{\mu}^{1-\delta}_{k,i-1}(\theta')L^{\delta}_k(\bm{\xi}_{k,i}|\theta')}
}
\label{eq:ASLinterm}
\end{equation}
where $0<\delta<1$ is a parameter that will be referred to as the {\em step-size}. In the second step, each agent $k$ aggregates all intermediate beliefs received from its neighbors into its updated belief vector $\bm{\mu}_{k,i}$ as 
\beq
\bm{\mu}_{k,i}(\theta)=\displaystyle{
	\frac{\exp\Big\{\sum_{\ell\in\mathcal{N}_k}a_{\ell k}\log\bm{\psi}_{k,i}(\theta)\Big\}}
	{\sum_{\theta'\in\Theta}
		\exp\Big\{\sum_{\ell\in\mathcal{N}_k}a_{\ell k}\log\bm{\psi}_{k,i}(\theta')\Big\}
	}
}.
\label{eq:ASLfinal}
\eeq

Different than classic social learning methods employed in~\cite{NedicTAC2017,Javidi,MattaSantosSayedICASSP2019,MattaBordignonSantosSayed2019}, we see that each agent performs the first step by modulating, through the convex weights $1-\delta$ and $\delta$, the relative weight assigned to the past and new information. In particular, relatively large values of $\delta$ give more importance to the new data, whereas small values of $\delta$ give more importance to the past beliefs. A similar form of convex combination appeared in the statistical literature for defining the Chernoff information~\cite{Chernoff1952}. 

As usual in the theory of adaptation and learning, the learning performance is characterized in the steady-state regime \cite{Sayed}. In steady state, the true hypothesis $\theta_0$ is kept constant over time, yielding:
\beq
\bm{\xi}_{k,i}\sim L_k(\xi|\theta_0),~~k=1,2,\ldots,N, ~~i=1,2,\ldots
\eeq
and that data $\{\bm{\xi}_{k,i}\}$ are independent and identically distributed (i.i.d.) over time. 

Let us define the log-likelihood ratio as
\beq
\bm{x}_{k,i}(\theta)\triangleq \log\left(\frac{L_k(\bm{\xi}_{k,i}|\theta_0)}{L_k(\bm{\xi}_{k,i}|\theta)}\right).
\label{eq:xkidefin}
\eeq

\begin{assumption}[Finiteness of KL divergences]\label{assum:integrable}
	For each $k=1,2,\dots,N$ and $\theta\neq\theta_0$:
	\beq
	d_k(\theta)\triangleq\E[\bm{x}_{k,i}(\theta)]<\infty.
	\label{eq:delldef}
	\eeq
	~\hfill$\square$
\end{assumption}
\vspace{-5pt}
To motivate cooperation among agents, we introduce the following identifiability assumption, which implies that the inference problem need not be locally identifiable.
\begin{assumption}[Global identifiability]\label{assum:globo}
	For each wrong hypothesis $\theta\neq\theta_0$, there exists at least {one agent $k_{\theta}$ that has strictly positive KL divergence, $d_{k_{\theta}}(\theta)>0$}.~\hfill$\square$
\end{assumption}

In order to characterize the learning performance, it is useful to introduce the logarithm of the ratio between the belief evaluated at $\theta_0$ and the belief evaluated at $\theta\neq\theta_0$:
\beq
\bm{\lambda}^{(\delta)}_{k,i}(\theta)\triangleq \log\left(\frac{\bm{\mu}_{k,i}(\theta_0)}{\bm{\mu}_{k,i}(\theta)}\right).
\eeq
When we omit the argument $\theta$ and write $\bm{\lambda}^{(\delta)}_{k,i}$, we are referring to the $(H-1)\times 1$ vector concatenating the log-belief ratios  $\bm{\lambda}^{(\delta)}_{k,i}(\theta)$ for $\theta\neq \thetatrue \in \Theta$. When we omit the subscript $i$ we are referring to a random variable characterized at the {\em steady state}, i.e., as $i\rightarrow\infty$. 
Thus, 
$
\bm{\lambda}^{(\delta)}_{k}(\theta) \textnormal{ and } \bm{\lambda}^{(\delta)}_{k}
$
are, respectively, the {\em steady-state} log-belief ratio evaluated at $\theta$, and the {\em steady-state} vector of log-belief ratios. 

\begin{remark}[Positive beliefs]
	\label{rem:muinit}
	In view of Assumption~\ref{assum:initbel} and the ASL algorithm \eqref{eq:ASLinterm}--\eqref{eq:ASLfinal}, we see that the belief $\bm{\mu}_{k,i}(\theta)$ remains nonzero for any $\theta$ across time. So, under stationary conditions, at any instant $i_0$, the belief vector $\bm{\mu}_{k,i_0}$ fulfills Assumption~\ref{assum:initbel}. This property allows performing the steady-state analysis from $i=0$ without losing generality. It is also relevant to avoid ill-defined log-belief ratios.~\hfill$\square$
\end{remark}

For each $i$, we can define the instantaneous decision of agent $k$ as corresponding to the hypothesis that maximizes the belief, which leads to the following error probability:
\begin{equation}
p^{(\delta)}_{k,i}=\P\left(\argmax_{\theta\in\Theta}\bm{\mu}_{k,i}(\theta)\neq\thetatrue\right)\stackrel{i\rightarrow  \infty}{\longrightarrow}p_k^{(\delta)},
\label{eq:errprob}
\end{equation}
where $p_k^{(\delta)}$ is the {\em steady-state} error probability.\footnote{{The existence of the limit in \eqref{eq:errprob} relies on the convergence proved in Theorem 1 (details omitted for space constraints).}}


\subsection{Network Average of Log-Likelihood Ratios}
First, a useful concept to introduce is the {\em network average} of log-likelihood ratios and its expectation, for all $\theta\neq\theta_0$:
\begin{IEEEeqnarray}{rCl}
\xnet(\theta)&=&\sum_{\ell=1}^N \pi_{\ell} \bm{x}_{\ell,i}(\theta),
\label{eq:avlik}\\
	\mnet(\theta)&\triangleq &\E[\xnet(\theta)]=\sum_{\ell=1}^N \pi_{\ell} d_{\ell}(\theta)
\label{eq:mnet}.
\end{IEEEeqnarray}
Second, if the log-likelihoods have finite variances\footnote{{Remarkably, the existence of second moments is not required in Theorems \ref{theor:steady} and \ref{theor:weaklaw}, and is used only in Theorem \ref{theor:CLT}.}}, we can compute the covariance {between $\bm{x}_{k,i}(\theta)$ and $\bm{x}_{k,i}(\theta')$ as}
\beq
\rho_{\ell}(\theta,\theta')=\E\left[
\Big(
\bm{x}_{\ell,i}(\theta)
-
d_{\ell}(\theta)
\Big)
\Big(
\bm{x}_{\ell,i}(\theta')
-
d_{\ell}(\theta')
\Big)
\right].
\eeq
Finally, if data are {\em independent across the agents}, the covariance between variables $\xnet(\theta)$ and $\xnet(\theta')$ is given as
{\begin{equation}
	\cnet(\theta,\theta')\triangleq
	\sum_{\ell=1}^N \pi^2_{\ell} \rho_{\ell}(\theta,\theta').
\label{eq:cnet}
\end{equation}}
\vspace{-10pt}
\section{Steady-State Analysis}
As seen in the introductory example, in the {\em adaptive} setting the belief will not converge (in the almost-sure sense) as $i\rightarrow\infty$. On the contrary, the belief of each agent will exhibit an {\em oscillatory} behavior: feature that enables adaptation. As we will see, because of this asymptotic {\em random} character, steady-state analysis is not trivial. The analysis of this random behavior is established in Theorem~\ref{theor:steady}.

Before stating the theorem, let us examine the evolution of the log-belief ratios. 
Manipulating \eqref{eq:ASLinterm} and \eqref{eq:ASLfinal} in the log domain, for every $\theta\neq\theta_0$ we have:
\begin{equation}
\bm{\lambda}^{(\delta)}_{k,i}(\theta)=(1-\delta)\sum_{\ell\in\mathcal{N}_k} a_{\ell k}\, \bm{\lambda}^{(\delta)}_{\ell,i-1}(\theta) 
+ 
\delta \sum_{\ell\in\mathcal{N}_k} a_{\ell k} \,\bm{x}_{\ell,i-1}(\theta).
\label{eq:mainASLrec}
\end{equation}
The recursion in \eqref{eq:mainASLrec} is in the form of a {\em diffusion} 
algorithm with {\em step-size} $\delta$ --- see, e.g., \cite{Sayed}. Developing the recursion in \eqref{eq:mainASLrec} we can write, for all $\theta\neq\theta_0$:
\begin{IEEEeqnarray}{rCl}
\bm{\lambda}^{(\delta)}_{k,i}(\theta)
&=&
(1-\delta)^i 
	\sum_{\ell=1}^N [A^i]_{\ell k} \bm{\lambda}_{k,0}(\theta)
\nonumber\\
\quad&+&\delta \sum_{m=0}^{i-1}\sum_{\ell=1}^N (1-\delta)^m [A^{m+1}]_{\ell k}\, \bm{x}_{\ell,i-m}(\theta).
\label{eq:withtransient}
\end{IEEEeqnarray}
Since the first term on the RHS of \eqref{eq:withtransient} vanishes as $i\rightarrow\infty$, for the steady-state analysis we can rewrite with slight abuse of notation:
\beq
\bm{\lambda}^{(\delta)}_{k,i}(\theta)=\delta \sum_{m=0}^{i-1}\sum_{\ell=1}^N (1-\delta)^m [A^{m+1}]_{\ell k}\, \bm{x}_{\ell,i-m}(\theta).
\label{eq:lambdarec}
\eeq

\begin{theorem}[Stability of log-belief ratios]
	\label{theor:steady}
	Let:
	\beq
	\bm{\lambda}^{(\delta)}_{k}(\theta)=
	\sum_{\ell=1}^N \delta \sum_{m=0}^{\infty} (1-\delta)^m [A^{m+1}]_{\ell k}\, \bm{x}_{\ell,m+1}(\theta),
	\label{eq:convseries}
	\eeq
	where the ordering of the summations in \eqref{eq:convseries} means that the $N$ inner series are all almost-surely convergent. Then, under Assumptions~\ref{assum:initbel} and~\ref{assum:integrable} we have that:
	\beq
	\boxed{
		\bm{\lambda}^{(\delta)}_{k,i}\stackrel{i\rightarrow\infty}{\rightsquigarrow} \bm{\lambda}^{(\delta)}_k
	}
	\eeq
	where $\rightsquigarrow$ indicates convergence in distribution.
	\QED
\end{theorem}
Theorem \ref{theor:steady} shows that, as long as the first moment of $\bm{x}_{k,i}$ exists, the statistical distribution of $\bm{\lambda}^{(\delta)}_{k,i}$ converges to the distribution of a stable (i.e., well-defined) random vector $\bm{\lambda}^{(\delta)}_k$ as $i\rightarrow\infty$. We remark that this does not imply that the partial sum in (15) will converge almost surely to \eqref{eq:convseries} as $i\rightarrow \infty$. The subtlety here is that while  
\begin{equation}
\widetilde{\bm{\lambda}}_{k,i}^{(\delta)}(\theta)\triangleq
\delta \sum_{m=0}^{i-1}\sum_{\ell=1}^N (1-\delta)^m [A^{m+1}]_{\ell k}\, \bm{x}_{\ell,m+1}(\theta)
\end{equation}
is almost-surely convergent (which can be deduced from part 1) of Lemma \ref{lem:mainlemma}, the summation in \eqref{eq:lambdarec} is not, due to the reversed ordering of the summands.

\section{Small-$\delta$ Analysis}
We will proceed with the asymptotic analysis of $\bm{\lambda}_{k}^{(\delta)}$ now in the regime of small $\delta$. As seen in~\cite{MattaSayedCoopGraphSP2018}, to deal with this asymptotic behavior in the adaptation context, we first introduce a steady-state vector $\bm{\lambda}^{(\delta)}_k$ that already embodies the effect of summing an infinite number of terms. Only then, we proceed to characterize the asymptotic behavior of the steady-state random vector $\bm{\lambda}^{(\delta)}_k$ as $\delta$ goes to zero. To support the results that follow, we will rely on Lemma~\ref{lem:mainlemma}, which can be found enunciated in Appendix \ref{ap:lem}.

\subsection{Consistent Social Learning}

\begin{theorem}[Consistency of ASL]
	\label{theor:weaklaw}
	Under Assumptions~\ref{assum:initbel} and~\ref{assum:integrable}, we have the following convergence:
	\beq
	\bm{\lambda}^{(\delta)}_k\stackrel{\delta\rightarrow 0}{\longrightarrow}\mnet~~\textnormal{ in probability}.
	\label{eq:wlawintermediate}
	\eeq
	Since under Assumption~\ref{assum:globo} all entries of $\mnet$ are strictly positive, Eq. \eqref{eq:wlawintermediate} implies that for all $\theta\neq\theta_0$:
	\beq
	\boxed{
		\lim_{\delta\rightarrow 0}p_k^{(\delta)}=0
	}
	\eeq i.e., each agent learns the truth as $\delta\rightarrow 0$.\QED
\end{theorem}

The result of Theorem~\ref{theor:weaklaw} relies on the weak law of small step-sizes proved in Lemma~\ref{lem:mainlemma}, part $3)$. This result requires the existence of the first moments $d_{\ell}(\theta)$, which is guaranteed by Assumption~\ref{assum:integrable}. Moreover, it requires that $\mnet(\theta)>0$ for all $\theta\neq\theta_0$, which is ensured by Assumption~\ref{assum:globo} and the strict-positivity of the Perron eigenvector.

\subsection{Normal Approximation for Small $\delta$}
Let us examine the behavior of the first two moments of the log-belief ratios. In view of Lemma~\ref{lem:mainlemma}, part $2)$, we conclude that the expectation of the steady-state random vector $\bm{\lambda}^{(\delta)}_k$ can be expressed as:
\begin{equation}
{\sf m}^{(\delta)}_k(\theta)\triangleq \E\left[\bm{\lambda}^{(\delta)}_k(\theta)\right]
=\mnet(\theta)+O(\delta),
\label{eq:mean}
\end{equation}
where $O(\delta)$ is a quantity such that the ratio $O(\delta)/\delta$ remains bounded as $\delta\rightarrow 0$.
Likewise, using part $4)$ of Lemma~\ref{lem:mainlemma}, we conclude that the covariance of the steady-state random vector $\bm{\lambda}^{(\delta)}_k$ results in: 
\begin{IEEEeqnarray}{rCl}
c^{(\delta)}_k(\theta,\theta')&\triangleq& \E\left[
\Big(\lambda^{(\delta)}_k(\theta)-{\sf m}^{(\delta)}_k(\theta)\Big)
\Big(\lambda^{(\delta)}_k(\theta')-{\sf m}^{(\delta)}_k(\theta')\Big)
\right]\nonumber\\
&=&
{\frac{\cnet(\theta,\theta')}{2}}\,\delta
+O(\delta^2).
\label{eq:covar}
\end{IEEEeqnarray}
Note that \eqref{eq:mean} and \eqref{eq:covar} can be rewritten in vector and matrix form, respectively as:
\beq
	{\sf m}^{(\delta)}_k=\mnet+O(\delta),\quad
	{\sf C}^{(\delta)}_k={\frac{\Cnet}{2}}\,\delta+O(\delta^2)
\label{eq:meancovarmat}
\eeq
where ${\sf C}^{(\delta)}_k=[c^{(\delta)}_k(\theta,\theta')]$ and $\Cnet=[\cnet(\theta,\theta')]$. 
The first equation in \eqref{eq:meancovarmat} shows that the expectation vector of the steady-state log-belief ratios, ${\sf m}^{(\delta)}_k$, approximates, for small $\delta$, the expectation vector of the average log-likelihood ratios, $\mnet$. Moreover, the second equation in \eqref{eq:meancovarmat} reveals that the covariance matrix of the steady-state log-belief ratios, ${\sf C}^{(\delta)}_k$, goes to zero as $\Cnet \,\delta/2$, where $\Cnet$ is the covariance matrix of the average log-likelihood ratios.

\begin{theorem}[Asymptotic normality]
	\label{theor:CLT}
	Assume that the data $\{\bm{\xi}_{k,i}\}$ are independent across the agents (recall that they are always assumed i.i.d. over time), and that the log-likelihood ratios have finite variance. 
	Then, under Assumptions~\ref{assum:initbel},~\ref{assum:integrable} and~\ref{assum:globo}, the following convergence in distribution holds:
	\beq
	\boxed{
		\frac{\bm{\lambda}_k^{(\delta)} - \mnet}{\sqrt{\delta}}
		\stackrel{\delta\rightarrow 0}{\rightsquigarrow} {\mathscr{G}\left(0,\frac{\Cnet}{2}\right)}
	}
	\label{eq:CLTstatement2}
	\eeq
where $\mathscr{G}(0,C)$ is a zero-mean multivariate Gaussian with covariance matrix equal to $C$.\QED
\end{theorem}
The result in Theorem \ref{theor:CLT} comes from Lemma \ref{lem:mainlemma}, part 5). As $\delta\rightarrow 0$, Theorem~\ref{theor:CLT} suggests the approximation:
\beq
\bm{\lambda}^{(\delta)}_k\approx{\mathscr{G}\left(\mnet, \frac{\Cnet}{2}\,\delta\right)}.
\label{eq:CLTfirstapp}
\eeq
{Using the expressions in \eqref{eq:mean} and \eqref{eq:covar} instead of the limiting $\mnet$ and $\Cnet\,\delta/2$, we get the alternative approximation:
\beq
\bm{\lambda}^{(\delta)}_k\approx\mathscr{G}\left({\sf m}^{(\delta)}_k, {\sf C}^{(\delta)}_k\right),
\label{eq:CLTsecondapp}
\eeq
which can capture different performance across agents, since ${\sf m}^{(\delta)}_k$ and ${\sf C}^{(\delta)}_k$ depend on $k$.}
\section{Simulation Results}
\label{sec:1}
\begin{figure}[htb]
	\centering
	\includegraphics[width=2.5in]{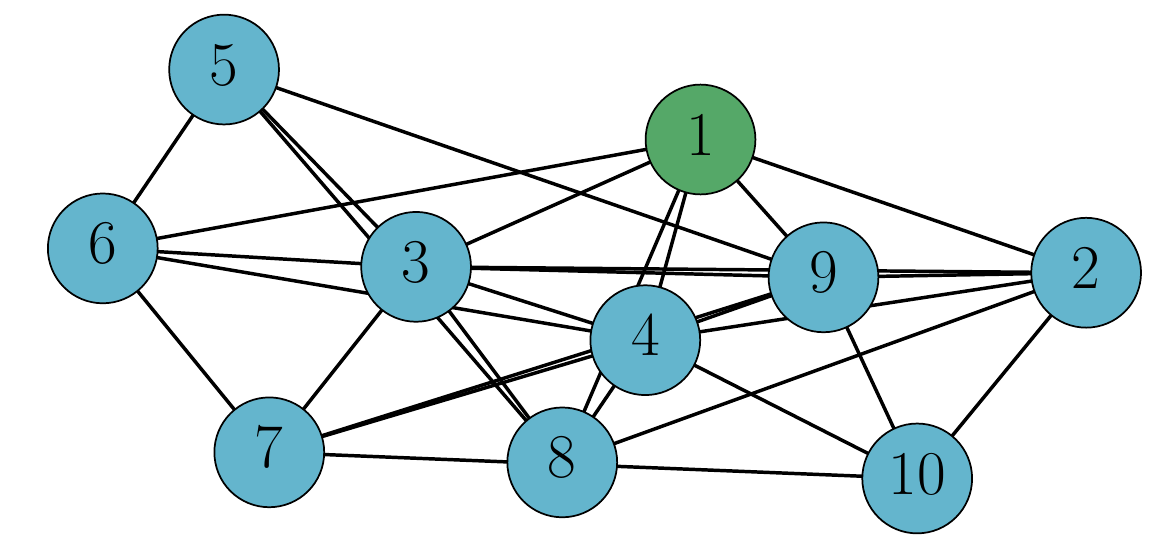}
	\caption{{Network topology with $10$ agents: agent $1$ is highlighted.}}
	\label{fig:network}
\end{figure}

{We consider the network topology displayed in Fig. \ref{fig:network} (additionally, we allow a self-loop for each agent).
The} combination matrix is designed using an averaging rule, resulting in a left-stochastic matrix \cite{Sayed}.
\vspace{-10pt}
\vspace{-5pt}

The network is faced with the following statistical learning problem.
We consider a family of Laplace likelihood functions with scale parameter equal to $1$, and with different expectations parametrized with $n\in\{1,2,3\}$ as follows:
\beq
f_n(\xi)=\frac{1}{2}\exp\left\{-|\xi-0.5 n|\right\}.
\label{eq:Gausspdf}
\eeq

We assume that the inference problem is {\em not locally identifiable} since we consider the setup in Table \ref{tab:id} for each agent's family of likelihood functions.
\vspace{-10pt}
\begin{table}[htbp]
	\def\arraystretch{1.3}%
	\caption{Identifiability setup for the network in Fig. \ref{fig:network}.}
	\begin{center}
		\begin{tabular}{|p{0.1\linewidth}|p{0.2\linewidth}|p{0.2\linewidth}|p{0.2\linewidth}|}
			\hline
			\multirow{2}{*}{\textbf{Agent} $k$}&\multicolumn{3}{|c|}{\textbf{Likelihood Function}: $L_k(\theta)$} \\
			\cline{2-4} 
			 &$\theta=1$ & $\theta=2$& $\theta=3$\\
			\hline
			$1-3$& $f_1(\xi)$& $f_1(\xi)$& $f_3(\xi)$ \\
			\hline
			$4-6$& $f_1(\xi)$& $f_3(\xi)$& $f_3(\xi)$ \\
			\hline
			$7-10$& $f_1(\xi)$& $f_2(\xi)$& $f_1(\xi)$ \\
			\hline
		\end{tabular}
		\label{tab:id}
	\end{center}
\vspace{-15pt}
\end{table}
\subsection{Consistency}
\begin{figure}[htb]
	\centering
	\includegraphics[width=3in]{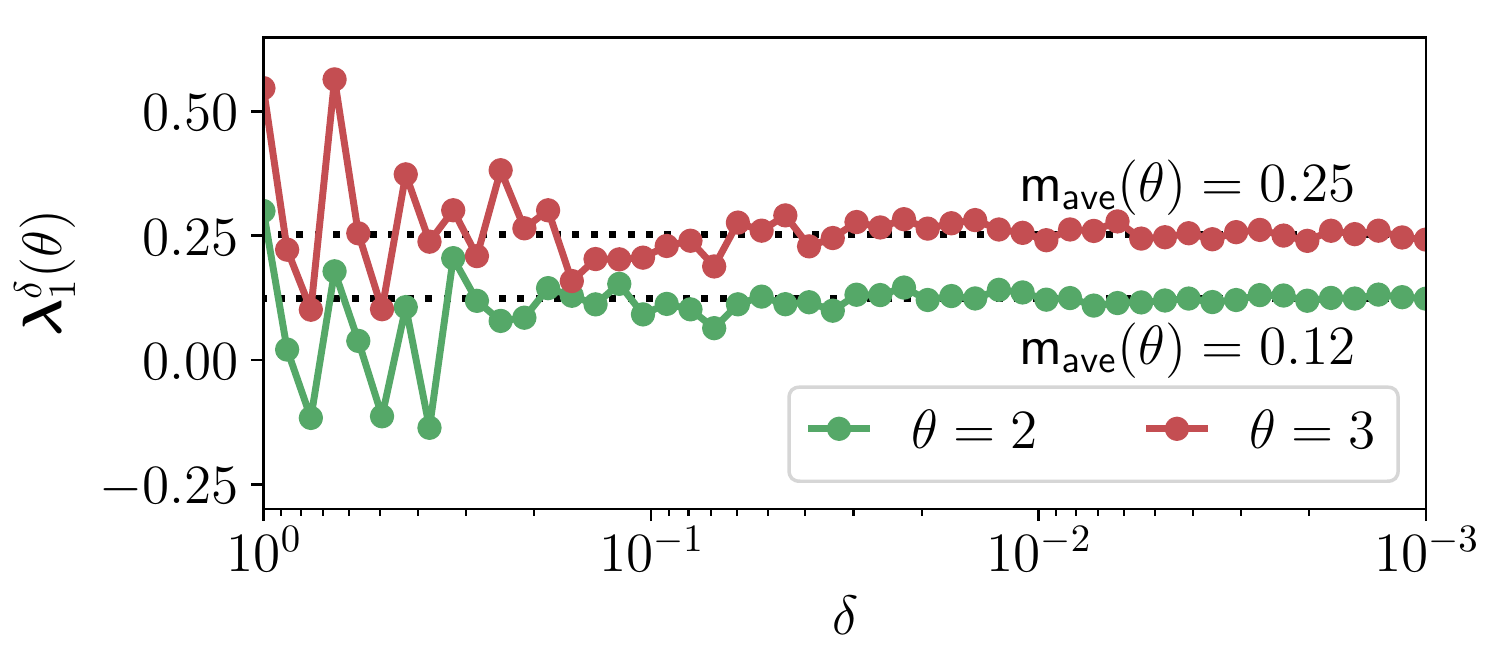}
	\caption{Evolution of steady-state log-belief ratios for agent $1$ as $\delta\rightarrow 0$.}
	\label{fig:theo1}
\end{figure}

We consider that all agents are running the ASL algorithm for a fixed $\thetatrue=1$ over $10000$ time samples (after which we consider that they achieved the steady state). From Theorem~\ref{theor:weaklaw}, we saw that as $\delta$ approaches zero, all agents $k$ are able to consistently learn --- see \eqref{eq:wlawintermediate}. 
To show this effect, for each value of $\delta$ (50 sample points in the interval $\delta\in[0.001,1]$ are taken), we consider a different realization of the observations. In Fig.~\ref{fig:theo1}, for agent $1$ and $\theta=2, 3$, we show how the log-belief ratios $\bm{\lambda}_1^{(\delta)}(\theta)$ behave for decreasing values of $\delta$. 
We see the effect of the {weak law of small step-sizes}, since the limiting log-belief ratios tend to concentrate around ${\sf m}_{\sf ave}$. 
\vspace{-5pt}
\vspace{-5pt}
\subsection{Asymptotic Normality}
\begin{figure}[htb]
	\centering
	\includegraphics[width=3.2in]{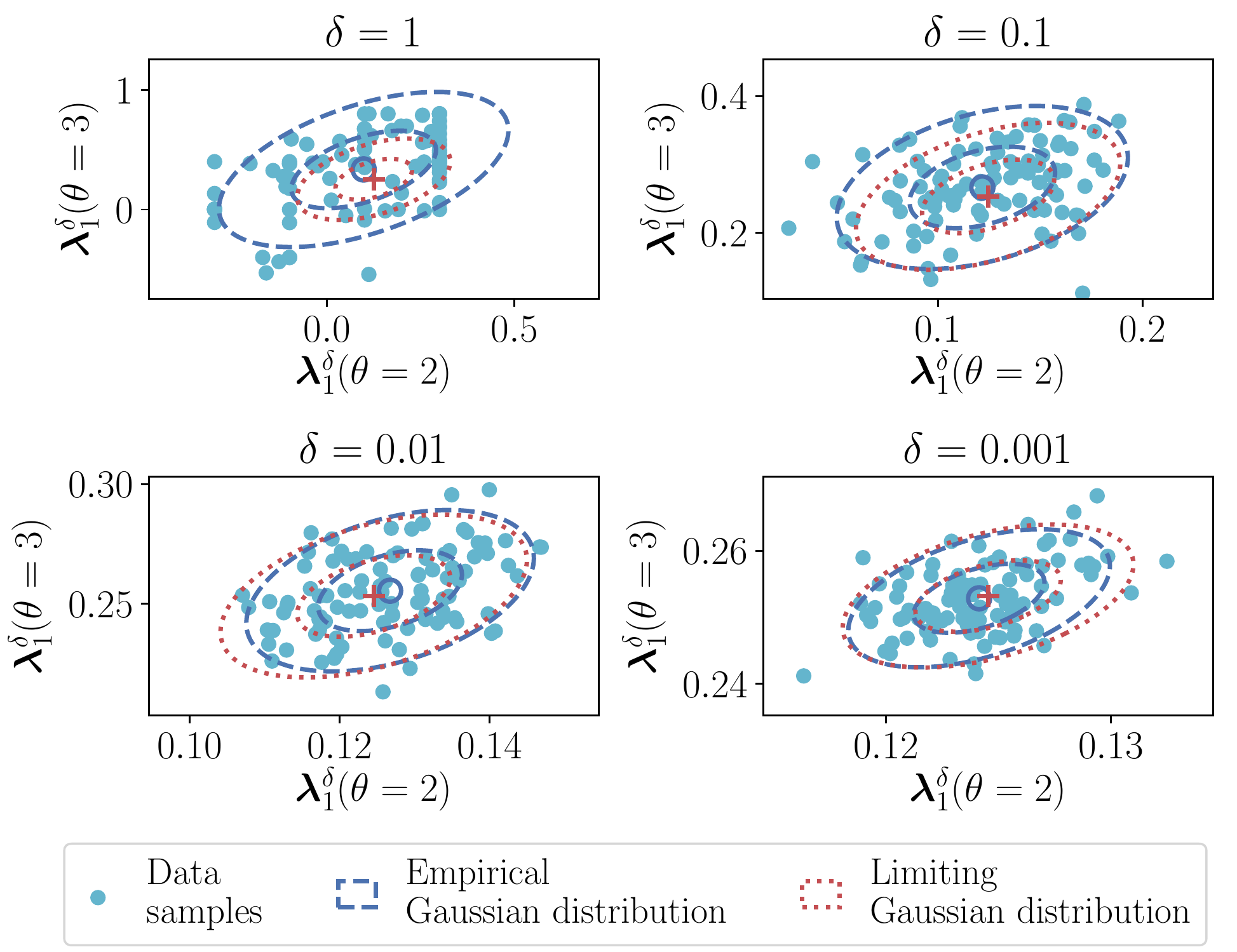}
	\caption{Distribution of data samples at steady state compared with the limiting and empirical Gaussian distributions for decreasing $\delta$.}
	\label{fig:theo2}
\end{figure}

From Theorem~\ref{theor:CLT}, we saw that we can approximate the steady-state log-belief ratios by a multivariate Gaussian, see Eqs. \eqref{eq:CLTfirstapp} and \eqref{eq:CLTsecondapp}. In Fig.~\ref{fig:theo2}, we display the log-belief ratios for instant $i=10000$. The experiment is repeated over $100$ Monte Carlo runs, such that we obtain $100$ realizations of the steady-state variable $\bm{\lambda}_k^{(\delta)}$. 
Moreover, we consider four decreasing values of $\delta$. 

In dashed blue lines we see the ellipses representing the confidence intervals relative to one and two standard deviations computed for the empirical Gaussian approximation seen in \eqref{eq:CLTsecondapp}: the smaller ellipse encompasses approximately $68\%$ of the samples whereas the larger ellipse encompasses $95\%$. In red dotted lines, we see the corresponding ellipses for the limiting theoretical Gaussian approximation seen in \eqref{eq:CLTfirstapp}, with the red cross indicating the limiting theoretical expectation ${\sf m}_{\sf ave}$. Note how as $\delta$ decreases, the ellipses tend to be smaller, which is in accordance with the scaling of the covariance matrices by $\delta$ in \eqref{eq:CLTfirstapp} and \eqref{eq:CLTsecondapp}, and the distributions tend to overlap, which is in accordance with the behavior predicted by Theorem~\ref{theor:CLT}.
\vspace{-5pt}
\section{Conclusion}
In this paper, we proposed the Adaptive Social Learning strategy as a way to address the significant inertia of the classic social learning to adapting. We have first characterized the behavior of the algorithm in steady state, by showing that the log-belief ratios converge to a stable random variable. Then, exploring the regime of small $\delta$, we could verify the algorithm's learning consistency and the limiting Gaussian behavior of the steady-state log-belief ratios.

\begin{appendices}
	\section{}\label{ap:lem}
	\begin{lemma}[Asymptotic properties of a useful random series]
		\label{lem:mainlemma}
		For $m=0,1,\ldots$, let $\{\bm{z}_m\}$ be a sequence of i.i.d. integrable random variables with ${\sf m}_z\triangleq\E\left[\bm{z}_m\right]$ and $
		{\sf m}^{\mathrm{abs}}_z\triangleq\E\left[|\bm{z}_m|\right]<\infty$. Let also $0<\delta<1$, and consider the following partial sums:
		\beq
		\bm{s}_i(\delta)=\delta
		\sum_{m=0}^{i}(1-\delta)^m \alpha_m \bm{z}_m,
		\label{eq:randomseries}
		\eeq
		where $0<\alpha_m<1$, with $\alpha_m$ converging to some value $\alpha$ and obeying the following upper bound for all $m$:
		\beq
		|\alpha_m - \alpha| \leq \kappa \beta^m,
		\label{eq:exprate}
		\eeq
		for some constant $\kappa>0$ and for some $0<\beta<1$.
		Then we have the following asymptotic properties.
		\begin{enumerate}
			\item 
			{\bf Steady-state stability}. The partial sums in \eqref{eq:randomseries} are almost-surely absolutely convergent, namely, we can define the (almost-surely) convergent series:
			\beqa
			\bm{s}^{\mathrm{abs}}(\delta)&\triangleq&\delta\sum_{m=0}^{\infty}(1-\delta)^m \alpha_m |\bm{z}_m|,
			\label{eq:limitdef1}\\
			\bm{s}(\delta)&\triangleq&\delta\sum_{m=0}^{\infty}(1-\delta)^m \alpha_m \bm{z}_m.
			\label{eq:limitdef2}
			\eeqa
			\item
			{\bf First moment}. The expectation of $\bm{s}(\delta)$ is:
			\beq
			\E[\bm{s}(\delta)]={\sf m}_z\delta\sum_{m=0}^{\infty}(1-\delta)^m \alpha_m=\alpha\,{\sf m}_z + O(\delta),
			\label{eq:expeclemma}
			\eeq
			where $O(\delta)$ is a quantity such that the ratio $O(\delta)/\delta$ remains bounded as $\delta\rightarrow 0$.
			\item
			{\bf Weak law of small step-sizes}. 
			The series $\bm{s}(\delta)$ converges to $\alpha\,{\sf m}_z$ in probability as $\delta\rightarrow 0$, namely, for all $\epsilon>0$ we have that:
			\beq
			\lim_{\delta\rightarrow 0}\P\left[|\bm{s}(\delta)-\alpha\,{\sf m}_z|>\epsilon\right]=0.
			\label{eq:weaklawequ}
			\eeq
			\item
			{\bf Second moment}. If
			$
			\sigma^2_z \triangleq\VAR[\bm{z}_m]<\infty,
			\label{eq:VARsingle}
			$
			then:
			\beqa
			\VAR[\bm{s}(\delta)]&=&
			\sigma^2_z\delta^2\sum_{m=0}^{\infty}(1-\delta)^{2m} \alpha^2_m\nonumber\\
			&=&\frac{\alpha^2\sigma^2_z}{2}\,\delta+O(\delta^2).
			\label{eq:VARlemma}
			\eeqa
			\item
			{\bf Asymptotic normality}. If $\bm{z}_m$ has finite variance $\sigma^2_z$, then the following convergence in distribution holds:
			\beq
			\frac{\bm{s}(\delta)-{\sf m}_z}{\sqrt{\delta}}\stackrel{\delta\rightarrow 0}{\rightsquigarrow} 
			\mathscr{G}\Big(0,\alpha^2\sigma^2_z/2\Big),
			\label{eq:CLTlemma}
			\eeq
			and, hence, $\bm{s}(\delta)$ is asymptotically normal as $\delta\rightarrow 0$.
		\end{enumerate}
	\end{lemma}

\end{appendices}

\end{document}